\documentclass[3p, twocolumn, number]{elsarticle}

\journal{Physics Letters A}
\usepackage{graphicx}
\usepackage{amssymb}
\usepackage{amsmath}

\begin{document}

\begin{frontmatter}

\title{Ultra-Fast Converging Path-Integral Approach for Rotating\\ Ideal Bose-Einstein Condensates}

\author[scl]{Antun Bala\v{z}\corref{cor}\fnref{fn1}}
\ead{antun@ipb.ac.rs}

\author[scl]{Ivana Vidanovi\'c\fnref{fn1}}

\author[scl]{Aleksandar Bogojevi\'c\fnref{fn1}}

\author[duisburg,potsdam]{Axel Pelster}

\address[scl]{Scientific Computing Laboratory, Institute of Physics Belgrade, Pregrevica 118, 11080 Belgrade, Serbia}
\address[duisburg]{Fachbereich Physik, Universit\" at Duisburg-Essen, Lotharstra\ss e 1, 47048 Duisburg, Germany}
\address[potsdam]{Universit\"at Potsdam, Campus Golm, Karl-Liebknecht-Strasse 24/25, 14476 Potsdam-Golm, Germany}

\cortext[cor]{Corresponding author.}
\fntext[fn1]{URL: http://www.scl.rs/}

\begin{abstract}
A recently developed efficient recursive approach for analytically calculating the short-time evolution of the one-particle propagator to extremely high orders is applied 
here for numerically studying the thermodynamical and dynamical properties of a rotating ideal Bose gas of $^{87}$Rb atoms in an anharmonic trap. At first, the one-particle 
energy spectrum of the system is obtained by diagonalizing the discretized short-time propagator. Using this, many-boson properties such as the condensation temperature, the 
ground-state occupancy, density profiles, and time-of-flight absorption pictures are calculated for varying rotation frequencies. The obtained results improve previous 
semiclassical calculations, in particular for smaller particle numbers. 
Furthermore, we find that typical time scales for a free expansion are increased by an order of magnitude for 
the delicate regime of both critical and overcritical rotation.
\end{abstract}


\end{frontmatter}

\section{Introduction}
\label{sec:introduction}

Bose-Einstein condensation (BEC) represents a macroscopic quantum phenomenon of recent broad research
interest \cite{pitaevskii, pethick, griffin, stoof}. Since its first experimental 
realization in 1995, it has been extensively studied experimentally, analytically, and numerically. The two main research directions are weakly-interacting dilute gases in 
magneto-optical traps and strongly-interacting quantum gases in optical lattices. The behavior of a Bose-Einstein condensate under rotation is essential for understanding many 
fundamental BEC phenomena. For instance, its response to rotation represents one of the seminal hallmarks of superfluidity \cite{fetter}. However, once a harmonically trapped 
Bose-Einstein condensate is 
rotated critically, i.e. the rotation frequency becomes so large that it compensates the radially confining harmonic frequency, the system turns out to be radially no 
longer trapped. In the absence of additional potential terms the condensate would start to expand perpendicular to the rotation axis. 
For an overcritical rotation, this expansion would even be accelerated by 
the presence of a residual 
centrifugal force. In order to reach experimentally this delicate regime of critical or overcritical rotation, Fetter suggested in Ref.~\cite{fetter2} to add an additional 
quartic term to the harmonic trap potential. Using a Gaussian laser beam propagating in the $z$-direction, this has been realized 
experimentally in Paris by Dalibard and co-workers for a 
BEC of $3\cdot 10^5$ atoms of $^{87}$Rb  \cite{dalibard1, bloch}. The resulting axially-symmetric trap with a small quartic anharmonicity in the $xy$-plane, seen by individual 
atoms, has the form
\begin{equation}
\label{eq:trap}
\hspace*{-1mm}
V_{\rm BEC}=\frac{M_\mathrm{Rb}}{2}(\omega_\perp^2-\Omega^2) r_\perp^2+\frac{M_\mathrm{Rb}}{2}\omega_z^2 z^2+\frac{k}{4}r_\perp^4\, ,
\end{equation}
with  the perpendicular radius $r_\perp =\sqrt{x^2+y^2}$, as well as the trap frequencies $\omega_\perp=2\pi\times 64.8$~Hz, $\omega_z=2\pi\times 11.0$~Hz, and the trap anharmonicity
$k=k_\mathrm{BEC}=2.6\times 10^{-11}~$Jm$^{-4}$. Furthermore,  the rotation frequency $\Omega$, which is measured in units of $\omega_\perp$, i.e. it is expressed by the ratio  $r = \Omega / \omega_\perp$, 
represents the tunable control parameter which could be experimentally increased up 
to $r=1.04$. This highest possible rotation frequency seems to coincide with an instability which follows from a Thomas-Fermi solution of the Gross-Pitaevskii equation
\cite{kling2}.

As long as we can ignore the presence of two-particle interactions and approximately describe the system with the ideal Bose gas, its many-particle properties in the 
grand-canonical ensemble are exclusively derivable from one-particle states. When considering the thermodynamic limit, usually the semiclassical approximation is applied, 
where the one-particle ground state $E_0$ is retained and treated quantum mechanically, while all one-particles states above $E_0$ are approximately treated as a continuum 
\cite{kling1}. This semiclassical approximation remains reasonable good irrespective of the rotation frequency $\Omega$ once the total particle number $N$ is large enough and 
the trap anharmonicity $k$ small enough. The latter condition implies that the underlying one-particle potential (\ref{eq:trap}) has a small curvature around its minimum. 
However, in this context the question arises for which system parameters such a semiclassical approximation is not sufficient for a precise description of BEC phenomena, as well 
as when it finally breaks down, requiring a full quantum mechanical treatment of the system.

In order to analyze this fundamental problem more quantitatively, it is mandatory to determine the one-particle energy eigenvalues and eigenfunctions
fully quantum mechanically. To this end we apply a recently 
developed ultra-fast converging path-integral approach for calculating the imaginary-time evolution amplitude 
of general non-relativistic many-body systems \cite{bogojevicprl, bogojevicprb, bogojevicpla, bogojevicplamb, balazpre}. In this approach a hierarchy of discretized effective 
actions is introduced where higher terms, compared to the naive action, substantially reduce errors in calculations of transition amplitudes. In particular, this allows
a systematic derivation of discretized effective actions of level $p$, which lead to a $1 / N^p$-convergence of discretized time-sliced transition
amplitudes within the continuum 
limit $N \to \infty$ of infinitely-many time slices. In addition, the improved convergence of transition amplitudes, calculated with the level $p$ effective action, can be 
used to construct higher-order analytic approximations for short-time propagators. 
Furthermore, the $N=1$ time-slice approximation turns out to be valid for short imaginary times and useful 
for small values of the inverse temperature in applications in quantum statistical physics. The recent Refs.~\cite{pqseeo1,pqseeo2} demonstrate that this path-integral approach 
allows to determine a huge number of both eigenvalues and eigenfunctions of quantum systems with very high accuracy using exact numerical diagonalization. 

In this Letter we show how this path-integral approach is applied for studying both global and local properties of fast-rotating Bose-Einstein condensates. To this end we proceed 
as follows: Sec.~\ref{sec:grand} briefly reviews the main ingredients of the path-integral approach in the general context of ideal Bose-Einstein condensates. Then we calculate 
in Sec.~\ref{sec:energy} a large number of energy eigenvalues and eigenfunctions for the anharmonic one-particle potential (\ref{eq:trap}). Afterwards, Sec.~\ref{sec:finite} 
discusses how a finite number of numerically available energy eigenvalues affects the results and how they can be improved by
introducing systematic semiclassical corrections. On the basis of 
this precise numerical one-particle information, Sec.~\ref{sec:global} 
studies global properties of a rotating condensate, for instance the condensation temperature $T_\mathrm{c}$ as a function 
of rotation frequency $\Omega$ and the ground-state occupancy $N_0/N$ as a function of temperature $T$. Finally,  Sec.~\ref{sec:local} is devoted to the calculation of local 
properties of the condensate, such as density profiles and time-of-flight absorption pictures. Sec.~\ref{sec:con} briefly summarizes the main results presented in this Letter.

\section{Many- Versus One-Particle Physics} 
\label{sec:grand}

At first we demonstrate that a precise numerical access to one-particle eigenstates allows the calculation of the condensation temperature and other thermodynamic properties of an 
ideal Bose gas. Afterwards, we briefly review details of the path-integral effective action approach for a numerical study of a $^{87}$Rb BEC in the anharmonic 
trap (\ref{eq:trap}).

\subsection{Ideal Bose Gas}

For high temperatures, the grand-canonical partition function of an ideal Bose gas is given by
\begin{equation}
\label{eq:Z}
{\cal Z}=\sum_\nu e^{-\beta(E_\nu-\mu N_\nu)}\, ,
\end{equation}
where $\nu$ enumerates all possible configurations of the system, $\beta = 1/k_B T$ represents the inverse temperature, and $\mu$ denotes the chemical potential. As the ideal bosons do not interact, the 
system energy $E_\nu$ can be expressed in terms of single-particle energy eigenvalues
\begin{equation}
E_\nu=\sum_n N_{\nu(n)}\, E_n\, ,
\end{equation}
where $n$ counts single-particle energy states, while $N_{\nu(n)}=0,1,2,\ldots$ and $E_n$ stand for the occupancy and the energy eigenvalue of level $n$, respectively. Correspondingly, the number of 
particles in the system reads
\begin{equation}
N_\nu=\sum_n N_{\nu(n)}\, .
\end{equation}
Thus, the grand-canonical free energy ${\cal F}=-(\ln {\cal Z})/\beta$ results to be
\begin{equation}
{\cal F}=\frac{1}{\beta}\sum_n\ln \left[ 1-e^{-\beta(E_n-\mu)} \right]\, ,
\end{equation}
where a subsequent Taylor expansion of the logarithm yields
\begin{equation}
\label{eq:F}
{\cal F}=-\frac{1}{\beta}\sum_{m=1}^\infty\frac{e^{m\beta\mu}}{m} Z_1(m\beta) \, ,
\end{equation}
usually denoted as the cumulant expansion. Thus, the many-body thermodynamic potential (\ref{eq:F}) of an ideal Bose gas is exclusively determined by single-particle states via the  
one-particle partition function:
\begin{equation}
\label{eq:Z1}
Z_1 (\beta)= \sum_{n} e^{- \beta E_n} \, .
\end{equation}

In principle, the above outlined exact calculation of the many-body free energy allows a 
full numerical description of ideal Bose gases, and can be also applied for studies of dilute Bose gases in the case 
when interactions are negligible. However, it becomes numerically very involved even for simple trapping potentials at low temperatures. In addition to this, the BEC phase transition is achieved only in 
the thermodynamic limit of an infinite number of atoms, thus making numerical studies of the condensation increasingly difficult. Usually, this problem is solved by fixing the chemical potential $\mu$ 
at the low temperatures of the condensate phase to the ground-state energy, i.e. by setting $\mu=E_0$, 
and to treat the ground state separately, by explicitly taking into account its macroscopic occupation $N_0$. Thus, for 
low enough temperatures the grand-canonical free energy (\ref{eq:F}) is modified to
\begin{eqnarray}
{\cal F}&=&-\frac{1}{\beta}\sum_{m=1}^\infty\frac{e^{m\beta\mu}}{m} \left[ Z_1(m\beta) - e^{- m \beta E_0} \right] \nonumber \\ 
&&+ N_0 ( E_0 - \mu)\, .
\end{eqnarray}
In order to avoid any double-counting, we have subtracted in the first line the contribution of the ground state within the one-particle partition function,
whereas the second line takes 
into account a possible macroscopic occupation of the ground state. The resulting total number of particles $N = - \partial {\cal F} / \partial \mu$ follows to be
\begin{equation}
\label{eq:N}
N=N_0+\sum_{m=1}^\infty e^{m\beta \mu}\left[ Z_1(m\beta)- e^{- m \beta E_0} \right]\, .
\end{equation}
This particle number equation serves different purposes in the respective phases. Within the gas phase, where the macroscopic occupation of the ground state vanishes, i.e.~we have $N_0=0$, Eq.~(\ref{eq:N})
determines the temperature dependence of the chemical potential $\mu$. On the other hand, 
within the BEC phase the chemical potential $\mu$ coincides with its minimal value, i.e.~the ground-state energy $E_0$,
so Eq.~(\ref{eq:N}) yields the temperature dependence of $N_0$. Therefore, the value of $\beta_\mathrm{c}=1/k_B T_\mathrm{c}$, which characterizes the boundary between both phases, follows from Eq.~(\ref{eq:N})
by setting $N_0=0$ and $\mu = E_0$:
\begin{equation}
\label{eq:Tc}
N=\sum_{m=1}^\infty \left[ e^{m\beta_\mathrm{c} E_0} Z_1(m\beta_\mathrm{c})-1 \right]\, .
\end{equation}
We conclude that, for a given number $N$ of ideal bosons, the condensation temperature can be exactly calculated only if both the single-particle ground-state energy $E_0$ and the full temperature 
dependence of the one-particle partition function (\ref{eq:Z1}) are known.

\subsection{Path-Integral Approach}

The recently developed path-integral approach \cite{bogojevicprl, bogojevicprb, bogojevicpla, bogojevicplamb, balazpre} allows an efficient and fast-converging numerical calculation of all one-particle 
properties of quantum systems \cite{pqseeo1, pqseeo2}. Note that this general numerical approach is suitable to treat arbitrarily-shaped trap potentials and is also applicable for exact studies of 
many-body problems. In this Letter we consider BEC experiments performed by Jean Dalibard's group \cite{dalibard1, bloch}, where a trapping potential with a quartic anharmonicity was realized. 
For this reason,  we will specialize our numerical approach in the following to such potentials in order to quantitatively analyze this seminal experiment.

The quantum statistical imaginary-time  transition amplitude $A(\mathbf a, \mathbf b;t)=\langle \mathbf b|e^{- t\hat{H}/\hbar} |\mathbf a \rangle $ 
follows in the continuum limit $N \to \infty$ of infinitely-many time slices
\begin{equation}
A(\mathbf a,\mathbf b;t)=\lim_{N \to \infty} A_N (\mathbf a,\mathbf b;t)\, .
\end{equation}
Here the time-sliced amplitude $A_N (\mathbf a, \mathbf b;t)$ 
is a product of short-time amplitudes corresponding to the introduced time steps, and is expressed as a multiple integral of the function $e^{-S_N/\hbar}$, 
where $S_N$ is usually called discretized (Euclidean) action \cite{kleinert}. 
For the simplest one-particle theory with the Euclidean Lagrangian $L=\dot{\mathbf q}^2 / 2 +V(\mathbf q)$, 
where the coordinates are rescaled so that the mass is equal to unity, the naive discretized 
action is given by
\begin{equation}
S_N=\sum_{n=0}^{N-1}\left[ \frac{\boldsymbol\delta_n^2}{2\varepsilon}+\varepsilon V(\bar{\mathbf q}_n)\right]\, ,
\end{equation}
with the abbreviations $\varepsilon=t/N$, $\boldsymbol\delta_n=\mathbf q_{n+1}-\mathbf q_n$, $\bar{\mathbf q}_n=(\mathbf q_{n+1}+\mathbf q_n)/2$ and boundary conditions $\mathbf q_0=\mathbf a$, 
$\mathbf q_N=\mathbf b$. Using the naive discretized action leads to the convergence of the 
transition amplitudes to the continuum as  slow as $1/N$.

Starting from the one-particle Schr\" odinger equation in $d$ spatial dimensions, we have derived a hierarchy of discretized effective actions $S_N^{(p)}$ with a systematically improved convergence to the 
continuum result. The general short-time amplitude is first written in the form
\begin{eqnarray}
&&\hspace*{-10mm}
A(\mathbf q_n,\mathbf q_{n+1};\varepsilon)\nonumber\\
&&=\frac{1}{(2\pi \hbar\varepsilon)^{d/2}}\, e^{-\frac{\boldsymbol\delta_n^2}{2 \hbar\varepsilon}-\frac{\varepsilon}{\hbar}W(\bar{\mathbf q}_n,\boldsymbol\delta_n;\varepsilon)}\, ,
\label{eq:Aqnqn1}
\end{eqnarray}
were $W(\bar{\mathbf q}_n,\boldsymbol\delta_n;\varepsilon)$ stands for the ideal discretized effective action, giving exact transition
amplitudes in any discretization. In fact, the above equation is valid for any propagation time 
$t$, not just for short times, and if we were able to calculate exactly the ideal effective potential $W$, we could use it directly to calculate the long-time 
transition amplitude $A(\mathbf a, \mathbf b;t)$. This is 
practically possible only for a limited number of exactly solvable models. However, we have derived a recursive approach which allows for an efficient analytical calculation 
of a high-order short-time expansion of 
the effective potential \cite{balazpre}. If we insert the expansion $W^{(p-1)}$ calculated to order $\varepsilon^{p-1}$ into Eq.~(\ref{eq:Aqnqn1}), we get the short-time amplitude
\begin{eqnarray}
&&\hspace*{-10mm}A^{(p)}(\mathbf q_n,\mathbf q_{n+1};\varepsilon)\nonumber\\
\label{eq:Ap}
&&= \frac{1}{(2\pi \hbar \varepsilon)^{d/2}}\, e^{-\frac{\boldsymbol\delta_n^2}{2 \hbar \varepsilon}-\frac{\varepsilon}{\hbar} W^{(p-1)}(\bar{\mathbf q}_n,\boldsymbol\delta_n;\varepsilon)}\, ,
\end{eqnarray}
which is designated by $p$ due to the fact that $W^{(p-1)}$ is multiplied by an additional factor of $\varepsilon$ in the exponent. This yields a result for the exponent which is
correct up to order $\varepsilon^p$, 
i.e. its error is proportional to $\varepsilon^{p+1}$. If we take into account the pre-factor $1/(2\pi\hbar\varepsilon)^{d/2}$, the above expression for the short time amplitude $A^{(p)}$ gives overall 
errors proportional to $\varepsilon^{p+1/2}$ in $d=1$. Consequently, in $d=2$ spatial dimensions, which is the case relevant here, the errors of short-time 
transition amplitudes would be of the order $\varepsilon^p$. However, if we consider the $N$ 
dependence of the above expression, the errors are always proportional to $1/N^{p+1}$, since the pre-factor $1/(2\pi\hbar\varepsilon)^{d/2}$ will be absorbed in the normalization of the amplitude 
$A(\mathbf a, \mathbf b;t)$. When used to calculate a long-time transition amplitude, the product of $N$ short-time amplitudes will subsequently lead to errors proportional to $1/N^p$, as demonstrated conclusively
earlier in Refs.~\cite{bogojevicprl, bogojevicprb, bogojevicpla, bogojevicplamb}.

Typical values of the inverse temperature $\beta$ in BEC experiments are quite small compared to the typical energy scale which is defined by the harmonic trap frequencies. For example, in the Paris
experiment \cite{dalibard1} the dimensionless value of $\hbar \beta \omega_\perp$ ranges between $10^{-3}$ and $10^{-1}$. Therefore, one can immediately use the above formula for the amplitude $A^{(p)}$ and 
calculate the corresponding one-particle
partition function by numerically integrating the diagonal amplitude $A(\mathbf a, \mathbf a, \beta)$ over the coordinate $\mathbf a$. For small enough $\beta$ the above formula converges rapidly, and 
the amplitudes can be calculated exactly for all practical purposes. We refer to this approximation as the $N=1$ approximation, since it corresponds to having only one time-slice in the standard definition 
of the transition amplitude in the path-integral formalism.

However, the direct use of this approach has several disadvantages. First of all, 
one still has to perform an integral over the diagonal coordinate $\mathbf a$ in order to calculate the partition function. Second, this 
has to be done repeatedly for each value of the inverse temperature $\beta$. And most importantly, one has also to extract the value of the ground-state energy in view of the particle
number equation (\ref{eq:N}).
In principle, this is done by studying the 
high-$\beta$ regime, where the short-time expansion (\ref{eq:Ap}) is not valid. Although this procedure works also for lower values of $\beta$ \cite{pla-danica}, it requires the numerical calculation of the 
one-particle partition function and a detailed study of its dependence on the inverse temperature in order to obtain the ground-state 
energy with sufficient precision. For this reason, the algorithm becomes numerically complex and difficult to use, especially in cases where the ground state is degenerate.

The approach based on a numerical diagonalization of the space-discretized evolution operator \cite{pqseeo1, pqseeo2} effectively resolves all of the above issues. It allows the precise numerical calculation 
of a large number of energy eigenvalues and eigenstates of the single-particle Hamiltonian. Once calculated in the low-$\beta$ regime, these data can be used to obtain the one-particle partition function 
for any value of the inverse temperature $\beta$ in a numerically inexpensive way. Calculations based on this approach do not have restrictions in the high-$\beta$ regime, where, in fact, they 
turn out to yield results 
with even better precision. In addition, the one-particle energy eigenfunctions obtained by the exact diagonalization will allow us to calculate local properties of Bose-Einstein condensates with very high 
accuracy.

\section{Numerical Calculation of Energy Eigenvalues and Eigenstates}
\label{sec:energy}

The most efficient method for calculating properties of few-body quantum systems is the direct diagonalization of the space-discretized propagator in imaginary time $\hat{U}(t)=\exp(-t \hat H / \hbar)$ 
\cite{pqseeo1, pqseeo2, sethia, sethiacpl1, sethiajcp, sethiacpl2}. Here $t$ represents the appropriately chosen propagation time, for which we can calculate the matrix elements of the evolution operator 
analytically within our path-integral approach in the $N=1$ approximation.

\begin{table}[!ht]
\begin{center}
\begin{tabular}{|c|c|c|}
\hline
&\multicolumn{2}{|c|}{$E_n/\hbar\omega_\perp$, $k=k_\mathrm{BEC}$}\\
\hline
$n$ & $r=0$&$r=1$\\
\hline\hline
$\, 0  \,  $ & $          1.0009731351803  $ &      0.1162667164134 \\\hline
$1               $ & $    2.0029165834022      $ & 0.2674689968905  \\\hline
$2               $ & $    2.0029165834022      $ & 0.2674689968905  \\\hline
$3               $ & $    3.0058275442161      $ & 0.4426927375269  \\\hline
$4               $ & $    3.0058275442161     $ &  0.4426927375270  \\\hline
$5               $ & $    3.0067964582067     $ &  0.4725275724941  \\\hline
$6               $ & $    4.0097032385903     $ &  0.6368178804983 \\\hline
$7               $ & $    4.0097032385903    $ &   0.6368178804984  \\\hline
$8               $ & $    4.0116368851078     $ &  0.6848142470356  \\\hline
$9               $ & $    4.0116368851078     $ &  0.6848142470357  \\\hline
\end{tabular}
\end{center}
\begin{center}
\begin{tabular}{|c|c|c|}
\hline
&\multicolumn{2}{|c|}{$E_n/\hbar\omega_\perp$, $k=10^3~k_\mathrm{BEC}$}\\
\hline
$n$ & $r=0$&$r=1$\\
\hline\hline
$\, 0  \,  $ & $          1.468486725893  $ &     1.162667164134 \\\hline
$1               $ & $    3.213056378201      $ & 2.674689968905  \\\hline
$2               $ & $    3.213056378201      $ & 2.674689968905  \\\hline
$3               $ & $    5.163819069871      $ & 4.426927375269  \\\hline
$4               $ & $    5.163819069871     $ &  4.426927375270  \\\hline
$5               $ & $    5.406908088225     $ &  4.725275724941  \\\hline
$6               $ & $    7.282930987460     $ &  6.368178804982 \\\hline
$7               $ & $    7.282930987460    $ &   6.368178804982  \\\hline
$8               $ & $    7.690584058915     $ &  6.848142470357  \\\hline
$9               $ & $    7.690584058915     $ &  6.848142470357  \\\hline
\end{tabular}
\end{center}
\caption{Lowest energy levels of the $xy$-part of the BEC potential (\ref{eq:trap}) for non-rotating ($r=0$) and critically rotating ($r=1$) condensate with the quartic anharmonicity  
$k=k_\mathrm{BEC}$ (top) and $k=10^3~k_\mathrm{BEC}$ (bottom). They are 
obtained by using level $p=21$ effective action with the discretization parameters of Table~\ref{tab:Emax}. The spacing $\Delta$ was always chosen 
so that $L/\Delta=100$, and the propagation time was $t=0.2$ for $k=k_\mathrm{BEC}$ and $t=0.05$ for $k=10^3~k_\mathrm{BEC}$. Errors are given by the precision of the last digit, typically $10^{-12}$ to $10^{-13}$, and are estimated by comparing the numerical results 
obtained with different discretization parameters.}
\label{tab:underandcrit}
\end{table}

\begin{table}[!t]
\begin{center}
\begin{tabular}{|c|c|c|}
\hline
&\multicolumn{2}{|c|}{$E_n/\hbar\omega_\perp$, $r=1.04$}\\
\hline
$n$ & $k_\mathrm{BEC}$&$10^3~k_\mathrm{BEC}$\\
\hline\hline
$\, 0  \,  $ &       -0.6617041825660 &  1.135693826206 \\\hline
$1               $ & -0.6465857464220 &  2.628129903790  \\\hline
$2               $ & -0.6465857464220  & 2.628129903790 \\\hline
$3               $ & -0.6032113415949  & 4.363876633929 \\\hline
$4               $ & -0.6032113415948  & 4.363876633929 \\\hline
$5               $ & -0.5349860004310  & 4.667653582963 \\\hline
$6               $ & -0.5349860004309  & 6.290444734007 \\\hline
$7               $ & -0.4451224795419  & 6.290444734007 \\\hline
$8               $ & -0.4451224795419  & 6.777210773169 \\\hline
$9               $ & -0.3362724309903  & 6.777210773169 \\\hline
\end{tabular}
\end{center}
\caption{Lowest energy levels of the $xy$-part of the BEC potential (\ref{eq:trap}) for over-critically rotating ($r=1.04$) condensate
with the quartic anharmonicity  $k=k_\mathrm{BEC}$ and $k=10^3~k_\mathrm{BEC}$ according to the same numerical procedure as in Table \ref{tab:underandcrit}.}
\label{tab:over}
\end{table}

On a given real-space grid defined by a set of coordinates $\mathbf x_{\mathbf j}=\mathbf j\Delta$, where $\Delta$ denotes the spacing, $\mathbf j \in [-L/\Delta , L/\Delta]^d$ is a $d$-dimensional vector of integers, and $L$ denotes the spatial cutoff, matrix elements of the  
propagator  are just short-time transition amplitudes
\begin{equation}
\label{eq:U}
U_{\mathbf i \mathbf j}(t)=\langle \mathbf i \Delta |\hat{U}(t)|\mathbf j\Delta\rangle \, .
\end{equation}
Note that throughout the paper we use dimensionless units, in which any energy is expressed in terms of $\hbar\omega_\perp$, while the length unit is the corresponding harmonic oscillator length $\sqrt{\hbar/M_\mathrm{Rb} \omega_\perp}$.

The eigenvectors $\psi_n(\mathbf j\Delta)$ of such a discretized evolution operator matrix correspond to the space-discretized eigenfunctions of the original Hamiltonian, and the corresponding eigenvalues $E_n$ 
are related to the eigenvalues of the single-particle Hamiltonian by $e^{-t E_n / \hbar}$. A detailed analysis of discretization errors of this method is given in Ref.~\cite{pqseeo1}. 
It  provides a practical 
algorithm for choosing optimal discretization parameters which minimizes the associated errors.

Another source of errors is related to the fact that for non-trivial potentials it is not possible to  calculate exactly evolution operator matrix elements. Higher-order effective actions minimize the 
errors for a given propagation time $t$ and allow an optimal choice of this parameter as well by taking into account other discretization errors, as was shown in Ref.~\cite{pqseeo2}. This approach is able 
to give very accurate energy eigenvalues even for moderate values of the propagation time $t$ of the order
$0.1$. Table~\ref{tab:underandcrit} presents the first several energy eigenvalues for the two-dimensional 
($xy$-) part of the BEC potential (\ref{eq:trap}) for the non-rotating case ($r=0$), as well as for the critically-rotating condensate ($r=1$). The top table gives the energy spectrum of the potential with 
the anharmonicity $k=k_\mathrm{BEC}$ used in the experiment \cite{dalibard1}, while the bottom table shows the spectrum for the much larger anharmonicity  $k=10^3~k_\mathrm{BEC}$. The degeneracies of numerically obtained eigenstates in all cases correspond to the expected structure of the spectrum, which can be deduced from the symmetry of the problem. In addition to this, the interesting case of critical rotation ($r=1$) allows a further verification of the numerical results. To this end we recall that the
energy eigenvalues of a pure quartic oscillator, to which $V_\mathrm{BEC}$ reduces in this case, are proportional to $k^{1/3}$ due to a spatial rescaling in the underlying Schr\"odinger equation. 
Therefore, we expect that the energy eigenvalues for $k=10^3~k_\mathrm{BEC}$ are precisely 10 times larger than the corresponding eigenvalues for $k=k_\mathrm{BEC}$. Comparing the rightmost columns in Table~\ref{tab:underandcrit} we see exactly this scaling. This  demonstrates conclusively that the 
presented method can be successfully applied also in this deeply non-perturbative parameter regime. Furthermore, Table~\ref{tab:over} gives the energy spectrum of an over-critically rotating ($r=1.04$) condensate, illustrating that the same approach can be used in this delicate regime as well.

\begin{figure}[!ht]
\centering
\includegraphics[width=7.8cm]{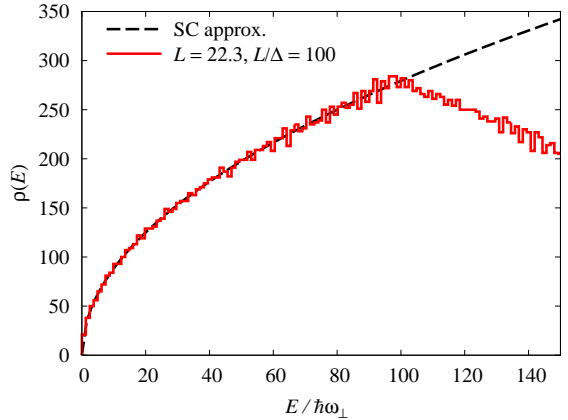}
\caption{Numerically calculated density of states for $xy$-part of the BEC potential with $k=k_\mathrm{BEC}$ for a critically rotating condensate, obtained by using level $p=21$ effective action. The discretization parameters are $L=22.3$, 
$L/\Delta=100$, and $t=0.2$. The dashed line is the corresponding semiclassical approximation for the density of states.}
\label{fig:DOS}
\end{figure}

With these results
single-particle partition functions $Z_1(\beta)$ can now be calculated according to Eq.~(\ref{eq:Z1}). This is especially suitable for the low-temperature regime, when higher energy levels give a 
negligible contribution. Although the above described approach is able to accurately give several thousands of energy eigenvalues, their number is always necessarily limited. This is easily seen from 
Fig.~\ref{fig:DOS}, where we compare the density of states for a critically rotating condensate with the corresponding semiclassical approximation for the
density of states \cite{pqseeo2}. This comparison allows to 
estimate the maximal reliable two-dimensional energy eigenvalue $E_\mathrm{max}$ which can be obtained numerically for a given set of discretization parameters. For example, from Fig.~\ref{fig:DOS} we can estimate 
$E_\mathrm{max}\approx 90$ for $r=1$ with the discretization parameters  $L=22.3$, $L/\Delta=100$, and $t=0.2$. Table~\ref{tab:Emax} gives estimates for the maximal reliable energy eigenvalue for the 
anharmonicites $k=k_\mathrm{BEC}$ and $k=10^3~k_\mathrm{BEC}$ for several values of rotation frequencies. These results are obtained from numerical calculations using the SPEEDUP 
codes \cite{speedup}. 
This table gives also an overview over those discretization parameters which were
used for a numerical diagonalization of the BEC potential (\ref{eq:trap}) in order to calculate both global and local properties of the 
condensate throughout the whole paper.

\begin{table}[!t]
\begin{center}
\begin{tabular}{|c|c|c|c|c|}
\hline
& \multicolumn{2}{|c|}{$k=k_\mathrm{BEC}$}&\multicolumn{2}{|c|}{$k=10^3~k_\mathrm{BEC}$}\\
\hline
$r$&$E_\mathrm{max}/\hbar\omega_\perp$ &$L$ &$E_\mathrm{max}/\hbar\omega_\perp$ &$L$\\
\hline\hline
$\, 0.0  \,  $ & 140 & 14.2 &190 & 3.90\\\hline
$\, 0.2  \,  $ & 140 & 14.4 & 190 & 3.90 \\\hline
$\, 0.4  \,  $ & 140 & 15.0 & 180 & 3.91 \\\hline
$\, 0.6  \,  $ & 140 & 16.3 & 180 & 3.92 \\\hline
$\, 0.8 \,  $ & 130 & 18.6 & 180 & 3.94\\\hline
$\, 1.0  \,  $ & 90 & 22.3 & 170 & 3.96 \\\hline
$\, 1.04  \,  $ & 90 & 23.2 & 170 & 3.96 \\\hline
\end{tabular}
\end{center}
\caption{Maximal reliable numerically calculated energy eigenvalue $E_\mathrm{max}$ of the $xy$-part of the BEC potential (\ref{eq:trap}) for different values of $r=\Omega/\omega_\perp$, estimated from
comparing the numerically obtained density of states $\rho(E)$ with the semiclassical approximation. The numerical diagonalization was done using  level $p=21$ effective action. The spacing $\Delta$ 
was always chosen so that $L/\Delta=100$, and the propagation time was  $t=0.2$ for $k=k_\mathrm{BEC}$ and $t=0.05$ for $k=10^3~k_\mathrm{BEC}$. The total number of reliable energy eigenstates is in all cases of the order of $10^4$.}
\label{tab:Emax}
\end{table}

In the low-temperature limit the finiteness of the 
number of known energy eigenstates does not present a problem. In fact, a precise knowledge of a large number of energy eigenvalues makes this approach a preferred method for 
a numerically exact treatment of low-temperature phenomena. On the other hand, the high-temperature regime, where thermal contributions of higher energy states play a significant role, is not treatable in 
the same way. This regime is usually not relevant for studies of BEC experiments, but we consider it for the sake of completeness. Furthermore,  we want to demonstrate that 
the effective action approach can even be successfully used for 
studying this parameter range. In the case that the temperature is high enough, 
so that effects of higher energy eigenstates cannot be neglected, the inverse temperature $\beta$ becomes a small parameter. Thus, it becomes possible to calculate 
numerically the single-particle partition function as a sum of diagonal amplitudes, i.e.
\begin{equation}
\label{eq:Z1trace}
Z_1(\beta)=\sum_{\mathbf j} U_{\mathbf j \mathbf j}(\beta)\Delta^d\, ,
\end{equation}
where $\Delta$ represents the spacing and, as before, the values of $\mathbf j$ are defined by $\mathbf j \in [-L/\Delta , L/\Delta]^d$, with the spatial cutoff $L$  chosen in such a way 
to ensure the localization of the evolution matrix 
within the interval $[-L, L]^d$. Note that this can be verified, since the evolution operator matrix elements are calculated using the analytic approximation (\ref{eq:Ap}) with level $p$ effective 
potential
\begin{equation}
\label{eq:Ukk}
U_{\mathbf j \mathbf j}^{(p)}(\beta)=\frac{1}{(2\pi\hbar^2\beta)^{d/2}}\, e^{-\beta W^{(p-1)}(\mathbf j\Delta, 0; \hbar\beta)}\,,
\end{equation}
where $\hbar\beta$ plays now the role of the imaginary time, and the influence of the kinetic term is not present since $U_{\mathbf j \mathbf j}$ is a diagonal transition amplitude ($\boldsymbol\delta=0$).

\section{Finite Number of Energy Eigenvalues and Semiclassical Corrections}
\label{sec:finite}

In the previous section we have described a numerical approach which is
capable of providing a large number of accurate energy eigenvalues for a general quantum system. Its major application is for few-body systems, 
where the complexity of the algorithm is very low. For instance, 
we are able to calculate typically $10^4$ energy eigenvalues for the considered BEC potential (\ref{eq:trap}). In this section we discuss in more 
detail how the finiteness of numerically available energy eigenstates affects the calculation of thermodynamic properties of Bose-Einstein condensates.

As outlined in Sec.~\ref{sec:grand}, the information on single-particle eigenvalues is sufficient for calculating the condensation temperature according to Eq.~(\ref{eq:Tc}). Below the condensation 
temperature, the ground-state occupancy follows from solving the equation
\begin{equation}
\label{eq:NN0}
N=N_0+\sum_{m=1}^\infty \left[e^{m\beta E_0}\, Z_1(m\beta)- 1\right]\, .
\end{equation}
In practical calculations, however, one is inevitably forced to restrict the sum over $m$ in the cumulant 
expansion (\ref{eq:NN0}) to some finite cutoff $M$, resulting in the following approximation for the number of thermal atoms
\begin{equation}
\label{eq:NN0M}
N-N_0=\sum_{m=1}^M\sum_{n=1}^\infty e^{-m\beta (E_n-E_0)}\, .
\end{equation}
Thus the ground-state occupancy $N_0$ depends for each particle number $N$ and temperature $T$ also on this cutoff $M$.
In particular, solving (\ref{eq:NN0M}) for the (inverse) condensation temperature, obtained by demanding $N_0=0$, will yield 
$\beta_\mathrm{c}(M)$ with an explicit dependence on $M$. The exact condensation temperature $\beta_\mathrm{c}$ is only obtained in the limit $M\to\infty$.

\begin{figure}[!t]
\centering
\includegraphics[width=7.8cm]{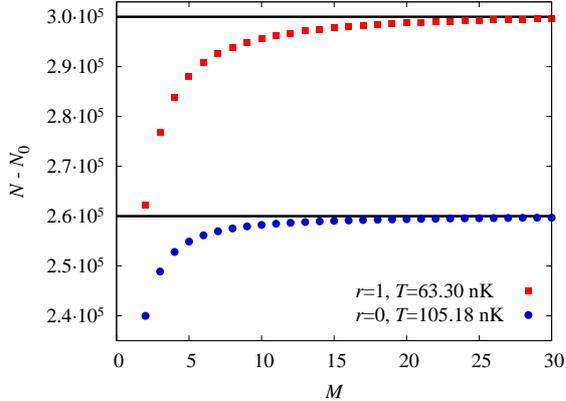}
\caption{Number of thermally excited atoms $N-N_0$ as a function of the cutoff $M$ in the cumulant expansion (\ref{eq:NN0M}). The results are given for a non-rotating condensate at $T=105.18$ nK and 
for a critically rotating condensate at $T=63.30$ nK. The results are obtained by level $p=21$ effective action, and all available
numerical eigenstates are used to calculate $N-N_0$. The discretization parameters 
were $L=14.2$ for $r=0$ and $L=22.3$ for $r=1$. In both cases the spacing was chosen according to $L/\Delta=100$, and the propagation time was $t=0.2$.}
\label{fig:N-T-M-noSC}
\end{figure}

Fig.~\ref{fig:N-T-M-noSC} illustrates the $M$-dependence resulting from Eq.~(\ref{eq:NN0M}) for both a non-rotating and a critically rotating condensate. As expected, the sum saturates for high values of 
$M$ to some finite number $N-N_0$. By tuning the temperature in such a way that the sum saturates at the desired value of the total number of atoms $N$ in the system, which implies $N_0=0$, 
one is, in principle, able to extract the condensation temperature $T_\mathrm{c}$.

\begin{figure}[!t]
\centering
\includegraphics[width=7.8cm]{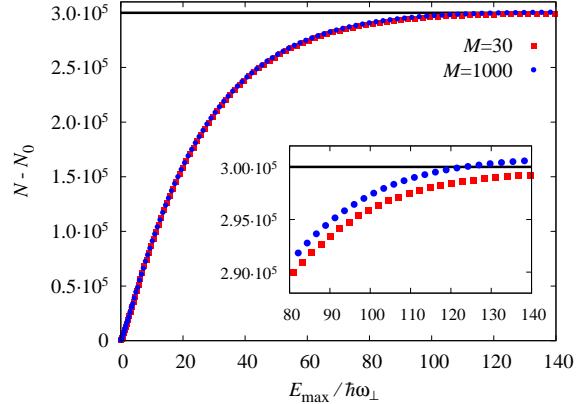}
\caption{Number of thermally excited atoms $N-N_0$ calculated as a function of the maximal available two-dimensional energy eigenvalue $E_\mathrm{max}$ at $T=63.30$~nK. The results are given 
for two different values of the cumulant cutoff $M$ for a critically rotating condensate, with the same parameters as in Fig.~\ref{fig:N-T-M-noSC}. The horizontal line corresponds to  the 
number of atoms $N=3\cdot 10^5$ in the experiment \cite{dalibard1}.}
\label{fig:N-T-Emax-noSC}
\end{figure}

Although the results in Fig.~\ref{fig:N-T-M-noSC} suggest that this approach can be applied straightforwardly, a closer look at the results for numerically calculated values of $N-N_0$ reveals several 
problems that have to be addressed. At first we have to investigate how the results depend
on the number of energy eigenstates used in the numerical calculation. Fig.~\ref{fig:N-T-Emax-noSC} gives this dependence for a
critically rotating condensate at its critical temperature $T_\mathrm{c}=63.30$ nK. 
We can see that the dependence on the maximal available two-dimensional energy eigenvalue $E_\mathrm{max}$ is quite significant. The inset of this figure reveals another problem: the value to which number 
$N_0-N$ saturates depends in addition 
on the cumulant cutoff $M$, as explained earlier. While the $M$-dependence can be dealt with by using a very large value of the cumulant cutoff in numerical calculations, the 
dependence on the maximal energy eigenvalue $E_\mathrm{max}$ must be eliminated by taking into account a proper semiclassical correction to the single-particle partition functions.

Namely, the finite number of energy eigenstates implies that the single-particle partition functions are only estimated by
\begin{equation}
Z_1(\beta)\approx \sum_{n=0}^{n_\mathrm{max}}e^{-\beta E_n}\, ,
\end{equation}
where $n_\mathrm{max}$ corresponds to the value $E_\mathrm{max}$ of the numerically available maximal 
energy eigenvalue. A semiclassical correction to this value, can be
calculated according to Ref.~\cite{kling1} as
\begin{eqnarray}
&&\hspace*{-1.2cm}\Delta Z_1(\beta, E_\mathrm{max})=\nonumber\\
&&\hspace*{-0.9cm}\int\frac{d^3\mathbf{x}\, d^3\mathbf{p}}{(2\pi\hbar)^3}\, e^{-\beta H(\mathbf{x}, \mathbf{p})}\, \Theta(H(\mathbf{x}, \mathbf{p})-E_\mathrm{max})\, ,
\label{semikling}
\end{eqnarray}
where $H(\mathbf{x}, \mathbf{p})$ represents the classical Hamiltonian of the system, while $\Theta$ denotes the Heaviside step-function.

For the trap potential (\ref{eq:trap})  we have in $z$-direction a pure harmonic potential which can be treated exactly. Therefore, we focus only on the two-dimensional problem in the 
$xy$-plane. In this case, 
the semiclassical correction for the single-particle partition function (\ref{semikling}) can be expressed in terms of the complementary error function:
\begin{eqnarray}
&&\hspace*{-1cm}\Delta Z_1^{(2)}(\beta, E_\mathrm{max})=\frac{1}{2\beta}\left\{  \frac{e^{-\beta E_\mathrm{max}}}{k} \left[-(1-r^2) \right. \right. \nonumber\\
&&\hspace*{-0.5cm}\left. +\sqrt{(1-r^2)^2+4kE_\mathrm{max}} \,\right]\, +\sqrt{\frac{\pi}{k\beta}} e^{\frac{\beta (1-r^2)^2}{4k}}\nonumber\\
&&\hspace*{-0.5cm}\left. \times  \mathrm{Erfc}\left(\sqrt{\beta E_\mathrm{max}+\frac{\beta (1-r^2)^2}{4k}}\right)\right\}\, .
\label{eq:2DSC}
\end{eqnarray}

\begin{figure}[!t]
\centering
\includegraphics[width=7.8cm]{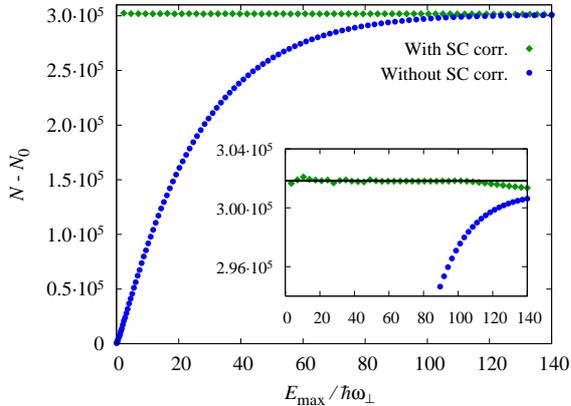}
\caption{Number of thermally excited atoms $N-N_0$ calculated as a function of $E_\mathrm{max}$ with and without semiclassical corrections, calculated with 
any large cumulant cutoff $M=10^4$ to eliminate the $M$-dependence. The results correspond to a critically rotating condensate with the same parameters as in Fig.~\ref{fig:N-T-M-noSC}. The horizontal line 
corresponds to $N=301 834$ which represents the exact value at $T_c=63.30$ nK.}
\label{fig:N-T-Emax-SC-M1e4}
\end{figure}

When this semiclassical correction is taken into account, the numerical results show almost no dependence on $E_\mathrm{max}$, as can be seen from Fig.~\ref{fig:N-T-Emax-SC-M1e4}. Here we have used an
excessively large value of the cumulant cutoff $M=10^4$ in order to completely eliminate any 
$M$ dependence. From the inset in this graph we also see that $E_\mathrm{max}$ must be chosen in accordance with the 
value estimated in the previous section for the maximal reliable energy eigenvalue obtained by numerical diagonalization. If we use a value $E_\mathrm{max}$ larger than this, we will be underestimating 
the higher part of the energy spectra, and obtain incorrect results. For a critically rotating condensate with the anharmonicity $k=k_\mathrm{BEC}$ the estimated value of $E_\mathrm{max}$ from 
Table~\ref{tab:Emax} is around 90 $\hbar\omega_\perp$, which agrees with the results from the inset of Fig.~\ref{fig:N-T-Emax-SC-M1e4}. If we use this value for $E_\mathrm{max}$ and calculate properties of 
the condensate using numerically obtained eigenstates below $E_\mathrm{max}$ with semiclassical corrections according to Eq.~(\ref{eq:2DSC}), we will obtain the exact results with very high accuracy.

\section{Global Properties of Rotating Bose-Einstein Condensates}
\label{sec:global}

In this section we will apply this approach to calculate different global properties of rotating Bose-Einstein condensates.

\subsection{Condensation Temperature}
\label{sec:Tc}

\begin{figure}[!t]
\centering
\includegraphics[width=7.8cm]{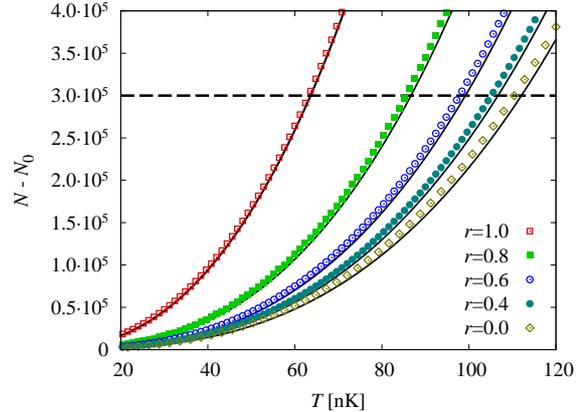}
\caption{Number of thermally excited atoms $N-N_0$ as a function of the temperature $T$ for different values of the rotation frequency and the quartic anharmonicity $k=k_\mathrm{BEC}$. The discretization 
parameters are given in Table~\ref{tab:Emax}, and the results are calculated by 
taking into account semiclassical corrections. The dashed line corresponds to  the number of atoms $N=3\cdot 10^5$ in 
the experiment \cite{dalibard1}. For comparison, the full lines depict the semiclassical results from Ref.~\cite{kling1}.}
\label{fig:N-T-T-knfac1}
\end{figure}

If we take into account semiclassical corrections as explained in the previous section, we can calculate, for instance, the condensation temperature of the condensate for different rotation 
frequencies. This implies that we have to 
find the temperature for which the number of thermal atoms $N-N_0$ 
saturates precisely at the total number of atoms $N$. In practice, this works the other way around: for a given condensation temperature $T_\mathrm{c}$ we 
numerically calculate the particle number in the system using Eq.~(\ref{eq:NN0M}), which gives the number of atoms in the system required for a condensation temperature to be equal to $T_\mathrm{c}$. This 
procedure is implemented in Fig.~\ref{fig:N-T-T-knfac1} for several values of the rotation frequency $\Omega$ in units of $r=\Omega/\omega_\perp$. For example, for $T_\mathrm{c}=63.14$ nK we see that the 
corresponding number of particles is $N=3\cdot 10^5$, which coincides with the value for a critically rotating condensate in the experiment of Dalibard and co-workers \cite{dalibard1}.

In principle, such a procedure is  only applicable for low-accuracy calculations of the critical temperature, 
since otherwise one has to use very large values of the cutoff $M$ which  
would practically slow-down numerical calculations. If one is interested in more precise results, a suitable $M$-dependence must be properly taken into account. In order to be able to efficiently extract
the correct value of $\beta_\mathrm{c}$, we will derive an analytical estimate of the asymptotic error $\Delta\beta_\mathrm{c}=\beta_\mathrm{c}-\beta_\mathrm{c}(M)$ which is
introduced by the cutoff $M$. Note that always $\Delta\beta_\mathrm{c}>0$, 
since $\beta_\mathrm{c}(M)<\beta_\mathrm{c}$ compensates the missing terms in the sum (\ref{eq:NN0M}).

If we insert $\beta_\mathrm{c}=\beta_\mathrm{c}(M)+\Delta\beta_\mathrm{c}$ into Eq.~(\ref{eq:NN0M}), the error $\Delta\beta_\mathrm{c}$ can considered to be
small for sufficiently large value of the cutoff $M$. By comparing 
Eq.~(\ref{eq:NN0M}) with the exact expression (\ref{eq:Tc}) we obtain
\begin{eqnarray}
&&\hspace*{-11mm}\sum_{m=M+1}^\infty \sum_{n=1}^\infty e^{-m\beta_\mathrm{c} (E_n-E_0)}\approx\nonumber\\
&&\hspace*{-8mm}\Delta\beta_\mathrm{c} \sum_{m=1}^M\sum_{n=1}^\infty m(E_n-E_0)\,e^{-m\beta_\mathrm{c} (E_n-E_0)} \, .
\end{eqnarray}
The term $m(E_n-E_0)$ within the sum can be obtained by setting $N_0=0$ and applying the partial derivative $\partial / \partial \beta_\mathrm{c}$ to Eq.~(\ref{eq:NN0M}):
\begin{eqnarray}
&&\hspace*{-11mm}-\Delta\beta_\mathrm{c}\frac{\partial N}{\partial\beta_\mathrm{c}} \approx\nonumber \\
&&\hspace*{-8mm}\left[1-\Delta\beta_\mathrm{c} \frac{\partial}{\partial\beta_\mathrm{c}}\right]\sum_{m=M+1}^\infty\sum_{n=1}^\infty e^{-m\beta_\mathrm{c} (E_n-E_0)}\, .
\label{eq:Deltaeq}
\end{eqnarray}
Note that the derivative of the particle number $N$ with respect to $\beta_\mathrm{c}$ is not equal to zero, since $N$ is here effectively defined by the sum (\ref{eq:Tc}). Therefore,  we have instead
\begin{equation}
\label{eq:Delta}
\hspace*{-1mm}\frac{\partial N}{\partial\beta_\mathrm{c}}
= -\sum_{m=1}^\infty\sum_{n=1}^\infty m\, (E_n-E_0)\, e^{-m\beta_\mathrm{c} (E_n-E_0)}\, .
\end{equation}
Clearly, the right-hand side is a negative quantity which does not depend on $M$. However, it does depend on $\beta_\mathrm{c}$ and the 
energy spectrum of the system. 

If the system is close to a $d$-dimensional harmonic oscillator, which is the case for the potential (\ref{eq:trap}) with the small anharmonicity relevant for the experiment, for large 
values of $M$ we have approximately
\begin{equation}
\hspace*{-1mm}
\sum_{m=M+1}^\infty\sum_{n=1}^\infty e^{-m\beta_\mathrm{c} (E_n-E_0)}\approx
d\, \frac{e^{-(M+1)\beta_\mathrm{c}\hbar\omega}}{1-e^{-\beta_\mathrm{c}\hbar\omega}}\, ,
\end{equation}
where $\omega$ denotes an effective harmonic frequency. For the case of a large anharmonicity, the effective frequency $\omega$ would depend on $k$, representing the harmonic expansion of the potential 
around its minimum. With such an estimate, Eq.~(\ref{eq:Deltaeq}) reduces to
\begin{eqnarray}
\label{eq:E1}
&&\hspace*{-12mm}
\Delta\beta_\mathrm{c}\approx -\frac{d}{1-e^{-\beta_\mathrm{c}\hbar\omega}}\times\nonumber\\
&&\hspace{-8mm} \frac{e^{-(M+1)\beta_\mathrm{c}\hbar\omega}}{\partial N / \partial \beta_\mathrm{c} 
+(M+1)\, e^{-(M+1)\beta_\mathrm{c}\hbar\omega}\frac{d\hbar\omega}{1-e^{-\beta_\mathrm{c}\hbar\omega}}}\, .
\end{eqnarray}
The term $(M+1)\, e^{-(M+1)\beta_\mathrm{c}\hbar\omega}$ in the denominator of the second factor can be neglected for large enough values of the cutoff $M$, yielding as a simplified version of the above expression:
\begin{eqnarray}
\label{eq:E2}
\Delta\beta_\mathrm{c}\approx - \frac{d\, e^{-(M+1)\beta_\mathrm{c}\hbar\omega}}{\partial N / \partial \beta_\mathrm{c} \, (1-e^{-\beta_\mathrm{c}\hbar\omega})}\, .
\end{eqnarray}

\begin{figure}[!t]
\centering
\includegraphics[width=7.8cm]{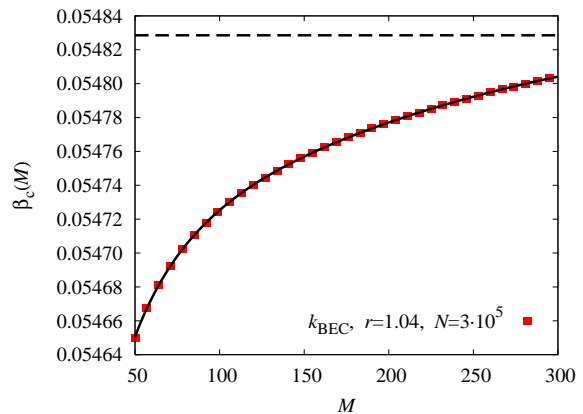}
\caption{Dependence of $\beta_\mathrm{c}$ on the cumulant cutoff $M$ for an over-critically ($r=1.04$) rotating condensate of $N=3\cdot 10^5$ atoms of $^{87}$Rb with the 
quartic anharmonicity of the trap $k=k_\mathrm{BEC}$. The discretization parameters are given in Table~\ref{tab:Emax}. The dashed line corresponds to a value of $\beta_\mathrm{c}$ obtained by fitting the numerical results to the function (\ref{eq:fit}), while the full line gives the fitted function $f(M)$.}
\label{fig:betac-E-SC-M}
\end{figure}

In order to use the derived estimates for $\Delta\beta_\mathrm{c}$, apparently one would already have to know the sought-after value of $\beta_\mathrm{c}$ as well as the difficult derivative 
$\partial N / \partial \beta_\mathrm{c}$. However, in practical applications this obstacle can be circumvented as follows. 
The expressions (\ref{eq:E1}) and (\ref{eq:E2}) can be used for fitting the numerical data 
for $\beta_\mathrm{c}(M)=\beta_\mathrm{c}-\Delta\beta_\mathrm{c}$, as is illustrated in Fig.~\ref{fig:betac-E-SC-M}. In this standard approach, all unknown values are fit parameters, obtained numerically by 
the least-square method. Note
 that not only $\beta_\mathrm{c}$ is obtained by such a fitting procedure, but also other parameters, such as $\partial N / \partial \beta_\mathrm{c}$, or the effective harmonic frequency $\omega$. The important point here 
is to capture the correct $M$-dependence, while all other parameters do not depend on it, so that they can be extracted by fitting. For example, in Fig.~\ref{fig:betac-E-SC-M} we have used the fitting 
function
\begin{eqnarray}
f(M)=\beta_\mathrm{c}-\frac{c_1e^{-c_2 (M+1)}}{1+c_3(M+1)\, e^{-c_4 (M+1)}}\, ,
\label{eq:fit}
\end{eqnarray}
which reproduces the numerical data quite accurately and gives high-precision results for the condensation temperature $T_\mathrm{c}$. The virtue of the derived estimates lies in the fact that they can be used to 
extract the information on the condensation temperature even for moderate values of $M$, when a saturation is not yet achieved. This substantially speeds up the numerical calculation of condensation 
temperatures, especially when it has to be done for different values of potential parameters, such as the frequency ratio $r=\Omega/\omega_\perp$.

\begin{figure}[!t]
\centering
\includegraphics[width=7.7cm]{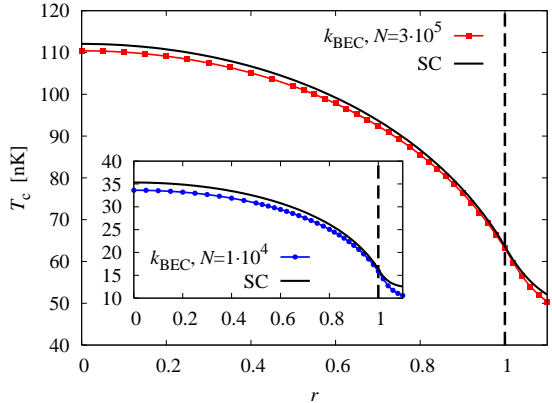}
\caption{The condensation temperature as a function of the rotation frequency for the condensate of $N=3\cdot 10^5$ and $N=1\cdot 10^4$ atoms of $^{87}$Rb, with the quartic anharmonicity of the trap 
$k=k_\mathrm{BEC}$. The discretization parameters are given in Table~\ref{tab:Emax}. The full lines correspond to the semiclassical approximation for $T_\mathrm{c}$ from Ref.~\cite{kling1}.}
\label{fig:Tc-r}
\end{figure}

Fig.~\ref{fig:Tc-r} summarizes the numerical results for the condensation temperature $T_\mathrm{c}$ for the anharmonicity $k=k_\mathrm{BEC}$ as well as the particle numbers
$N=3\cdot 10^5$ and $N=1\cdot 10^4$.  If we compare the obtained 
numerical results with the semiclassical approximation  from Ref.~\cite{kling1}, we see that the agreement turns out to be relatively good 
for the undercritical regime, but it becomes worse for an overcritical rotation of the 
condensate. After presenting results for the ground-state occupancy, which were obtained from this approach in the next section, 
we will compare our numerically exact results with the semiclassical approximation in more 
detail, and identify the parameter ranges where a full numerical treatment becomes necessary.

\subsection{Ground-State Occupancy}
\label{sec:N0N}

The ground-state occupancy is the next important global property of Bose-Einstein condensates we will look into. Below the condensation temperature a non-trivial fraction of atoms is in the ground 
state, thus yielding a macroscopic value of the occupancy ratio $N_0/N$.

\begin{figure}[!t]
\centering
\includegraphics[width=7.8cm]{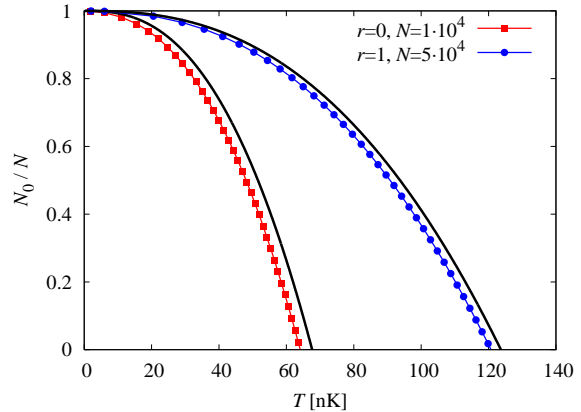}
\caption{Ground-state occupancy $N_0/N$ as a function of the temperature $T$ for non-rotating and critically rotating condensate for different values of the total number of atoms. The quartic anharmonicity 
is $k=10^3~k_\mathrm{BEC}$, and the discretization parameters are given in Table~\ref{tab:Emax}. The full lines depict the semiclassical results from Ref.~\cite{kling1}.}
\label{fig:N0N-T-knfac1e3}
\end{figure}

Using the same approach as above, we can calculate the ground-state occupancy from Eq.~(\ref{eq:NN0}). After determining the ground-state energy $E_0$ from an 
exact diagonalization of the evolution operator, we obtain
\begin{equation}
\label{eq:Nnew}
\frac{N_0}{N}=1-\frac{1}{N}\sum_{m=1}^\infty \left[ e^{m\beta E_0}{\cal Z}_1 (m\beta)-1 \right ]\, .
\end{equation}
In order to calculate $N_0/N$, we need the full single-particle energy spectrum. For low temperatures, the large number of energy eigenstates obtained within the
exact diagonalization is sufficient. In the 
sum of Eq.~(\ref{eq:Nnew})
we have again to introduce a cutoff $M$ and to eliminate it by applying the methods described in previous sections. To this end one uses either
a very large value for the cutoff or one derives the appropriate finite 
correction term, and fits the results to the derived function.

Fig.~\ref{fig:N0N-T-knfac1e3}  presents numerical results for the ground-state occupancy of the condensate. The quartic anharmonicity of the trap is chosen to be
$k=10^3~k_\mathrm{BEC}$, and the resuts are given for 
the nonrotating case with the total number of atoms $N=1\cdot 10^4$ and for critically rotating codensate with $N=5\cdot 10^4$ atoms. A comparison with the semiclassical results derived in 
Ref.~\cite{kling1} shows that the deviations increase with larger temperatures and smaller particle numbers.

\subsection{Comparison with Semiclassical Approximation}

\begin{figure}[!t]
\centering
\includegraphics[width=7.8cm]{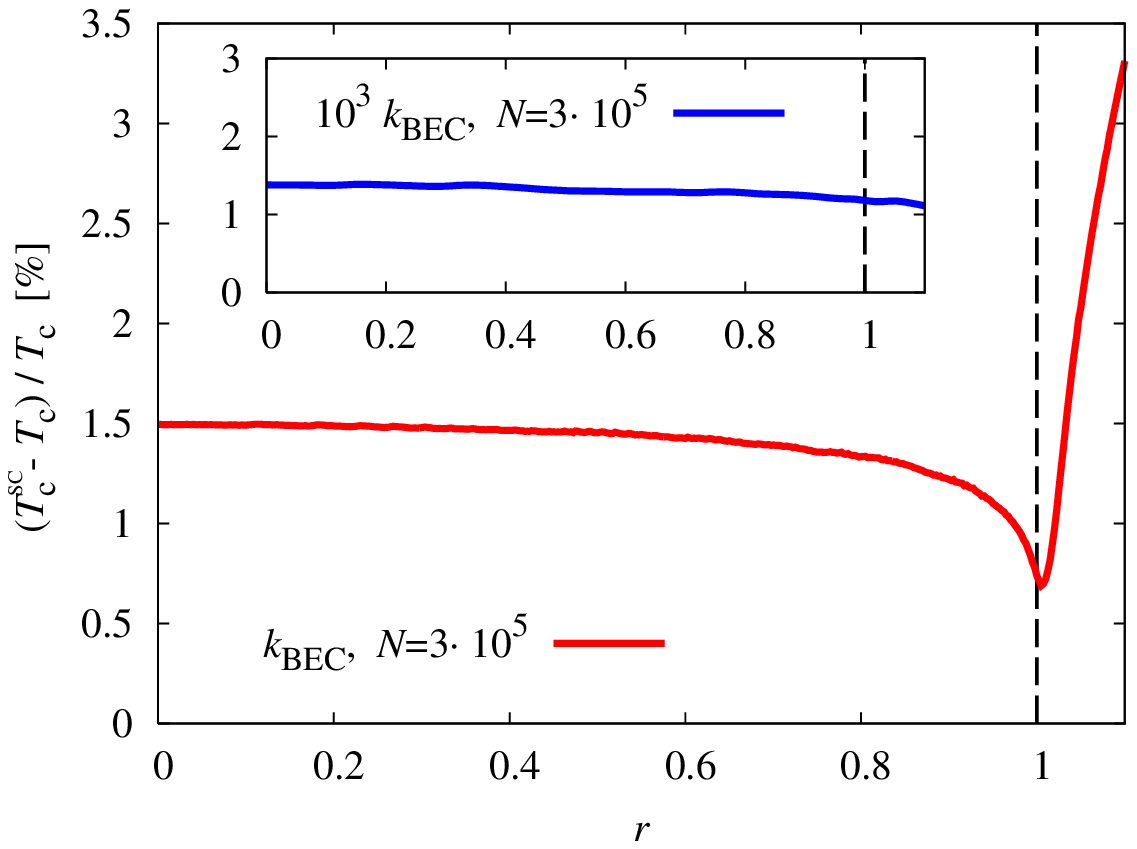}
\includegraphics[width=7.8cm]{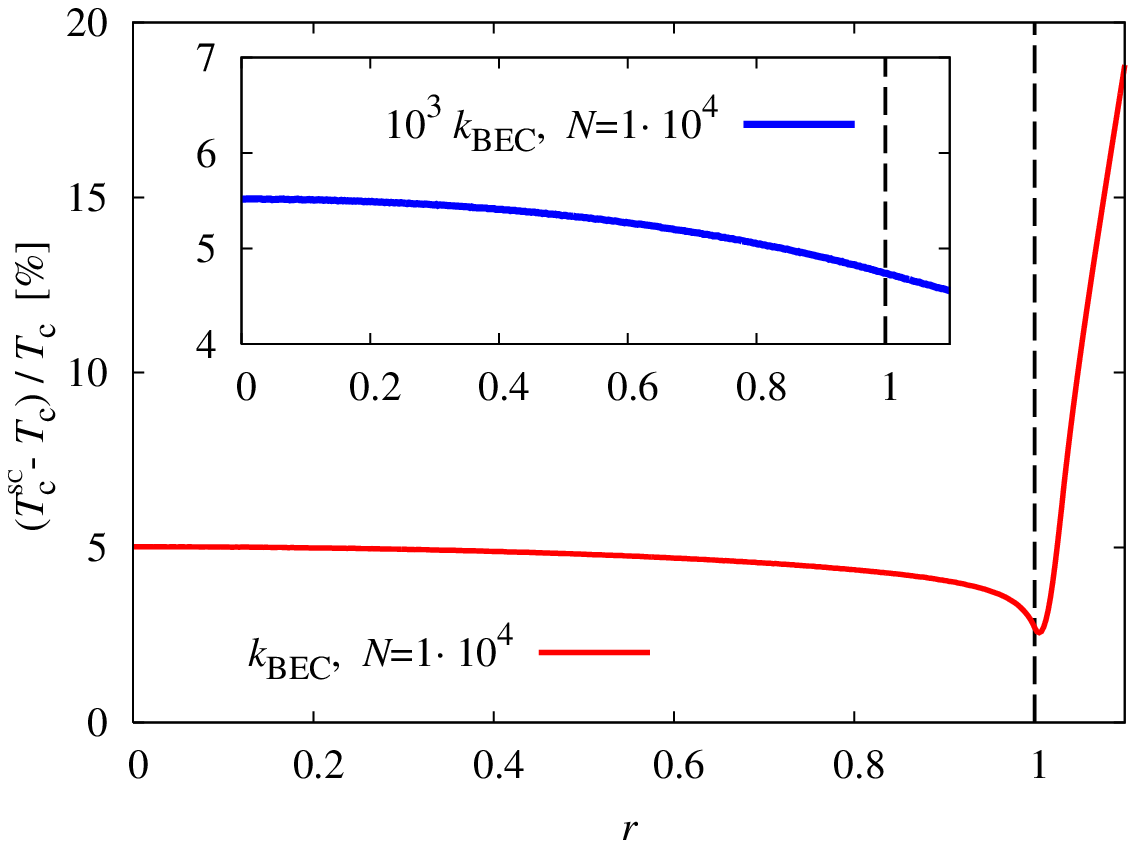}
\caption{Relative error of semiclassical results for the condensation temperature \cite{kling1} as a function of the rotation frequency $\Omega$  in units of
$r=\Omega/\omega_\perp$ for $N=3\cdot 10^5$ 
(top) and $N=1\cdot 10^4$ (bottom). The quartic anharmonicity is $k=k_\mathrm{BEC}$, and the discretization parameters are given in Table~\ref{tab:Emax}. The insets in both plots give the corresponding 
results for the large anharmonicity $k=10^3~k_\mathrm{BEC}$.}
\label{fig:Tc-E-SC-SC}
\end{figure}

Fig.~\ref{fig:Tc-E-SC-SC} depicts the errors of the
semiclassically calculated condensation temperature, where the exact values are obtained by using the presented numerical approach. As we can see, the agreement 
is relatively good for large particle numbers and small anharmonicity if the condensate rotates under-critically. The error in this case is of the order of 1~\% to 1.5~\%, and 
turns out to be minimal for a critical 
rotation. However, the error significantly increases for an overcritical rotation up to almost 3.5~\% for $r=1.1$. Therefore, while the
semiclassical approximation is acceptable for undercritical rotation, in the 
overcritical regime a numerical treatment becomes necessary. This is even more pronounced if we decrease the particle number to $10^4$, which is quite typical for many BEC experiments. In that case, 
semiclassical results already have an error of the order of 20~\%. For large anharmonicity the rotation effect is not so important, as we can see from insets on both graphs in Fig.~\ref{fig:Tc-E-SC-SC}. 
However, increasing the particle number to $10^4$ makes a numerical treatment indispensable, since the errors of the semiclassical results amount up to 5~\%.

\section{Local Properties of Rotating Bose-Einstein Condensates}
\label{sec:local}

Local properties of ultra-cold quantum gases are ubiquitously
used to observe and study the phenomenon of Bose-Einstein condensation. The prominent peak in time-of-flight absorption pictures, which appears suddenly 
when the temperature is decreased below $T_\mathrm{c}$, is a clear signature for the occurrence of 
a BEC phase transition. It is experimentally used to measure the thermodynamic properties of the condensate. In 
this section we will show how the presented numerical approach can be applied to calculate both the density profiles and the time-of-flight absorption imaging profiles.

\subsection{Density Profiles}

The two-point propagator $\rho(\mathbf x_1,\mathbf x_2)=\langle\hat\Psi^\dagger(\mathbf x_1) \hat\Psi(\mathbf x_2)\rangle$ defines via its diagonal element, i.e. 
$n(\mathbf x)=\rho(\mathbf x,\mathbf x)$, the density profile of atoms in a trap.
For the ideal Bose gas, the density profile can be written as
\begin{equation}
n(\mathbf x)=N_0|\psi_0(\mathbf x)|^2+\sum_{n\geq1}N_n |\psi_n(\mathbf x)|^2\, ,
\label{eq:npsi}
\end{equation}
where the second term represents the thermal contribution to the density profile. Furthermore, the quantities $\psi_n$ represent single-particle eigenstates, while the occupancies $N_n$ with $n\geq 1$
are given by the Bose-Einstein distribution
\begin{equation}
N_n=\frac{1}{e^{\beta (E_n-E_0)}-1}\, .
\end{equation}

\begin{figure}[!t]
\centering
\includegraphics[width=6.7cm]{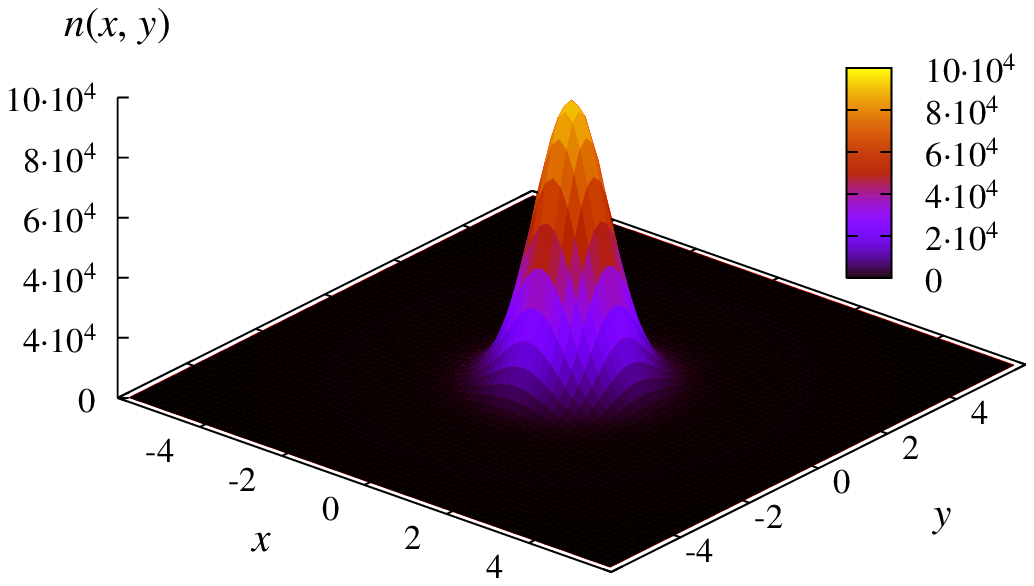}
\includegraphics[width=6.7cm]{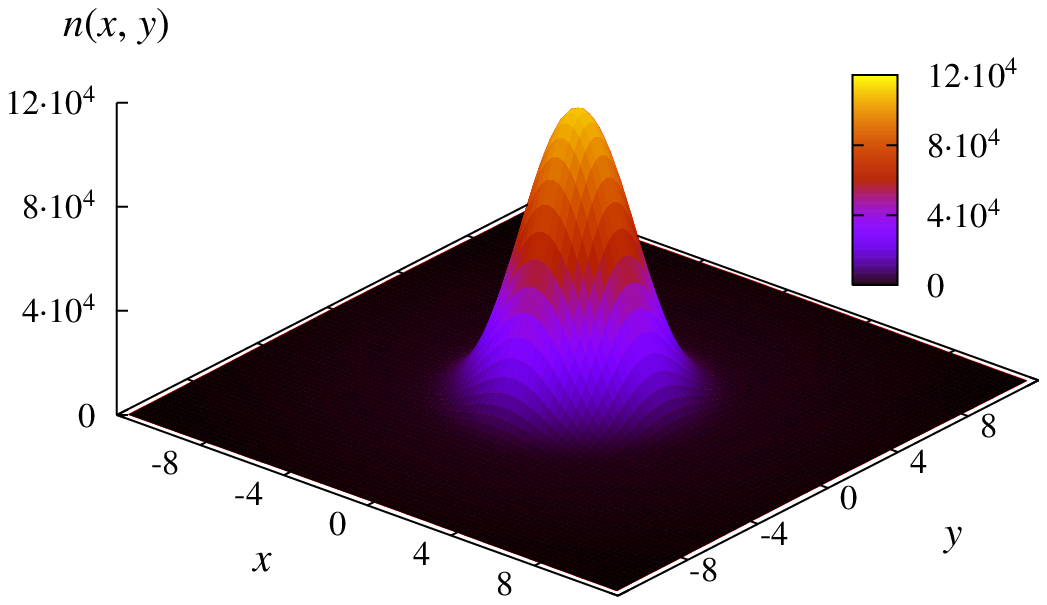}
\caption{Density profile in $xy$-plane for a non-rotating (top) and a
critically rotating (bottom) condensate of $N=3\cdot 10^5$ atoms of $^{87}$Rb with the anharmonicity $k=k_\mathrm{BEC}$ at $T=30$ nK. The 
dimensionless unit length on both graphs corresponds to 1.34~$\mu$m, i.e.~the linear size of the profile is approximately 16.1~$\mu$m (top) and 32.2~$\mu$m (bottom). The discretization parameters are given 
in Table~\ref{tab:Emax}.}
\label{fig:n-knfac1-T30}
\end{figure}

Having at our disposal numerically calculated energy eigenvalues and eigenfunctions, we can calculate the density profile of the condensate. In order to do so, we first have to obtain the ground-state 
occupancy number $N_0$ using the approach described in the previous section. Once this is done, Eq.~(\ref{eq:npsi}) 
allows to calculate the density profile. In view of a comparison with absorption imaging, which always produces
two-dimensional profiles, we have to integrate our numerically determined three-dimensional particle density $n(\mathbf x)$ along the imaging axis. Fig.~\ref{fig:n-knfac1-T30} 
presents typical results for the resulting density profiles of Bose-Einstein condensates for both 
the non-rotating and the critically-rotating case. Obviously, a rotation of the condensate leads to an effective
spreading due to the appearance of a centrifugal potential.

Although this approach is sufficient for treating the low-temperature regime, where the condensate is present, we  emphasize that the same method can also be used to deal with the thermal regime, when the 
temperature is increased beyond $T_\mathrm{c}$. For even higher temperatures, when the number of energy eigenstates, that need to be taken into account, exceeds the number of numerically accessible eigenstates, the presented 
approach can be extended in a similar way as the partition function was calculated previously as a sum of diagonal transition amplitudes. Using the cumulant expansion of occupancies and the spectral 
decomposition of thermal transition amplitudes, the density profile can be written for high enough temperatures as
\begin{eqnarray}
&&\hspace*{-12mm}n(\mathbf x)=N_0|\psi_0(\mathbf x)|^2\nonumber\\
&&\hspace*{-7mm}+ \sum_{m\geq1}\left[e^{m\beta E_0}A(\mathbf x, 0; \mathbf x, m\beta\hbar) - |\psi_0(\mathbf x)|^2\right]\, .
\label{eq:nA}
\end{eqnarray}
Here $A(\mathbf x, 0; \mathbf x, m\beta\hbar)$ represents the imaginary-time amplitude for a single-particle transition from the position $\mathbf x$ in the initial imaginary time $t=0$ to the position $\mathbf x$ 
in the final imaginary time $t=m\beta\hbar$.

While both definitions (\ref{eq:npsi}) and (\ref{eq:nA}) are mathematically equivalent in the case when one is able to calculate infinitely many energy eigenstates and amplitudes for an arbitrary  
propagation time, 
the first one is more suitable for low temperatures, when the number of relevant energy eigenstates is moderate, and the second one is suitable for high temperatures, when the 
imaginary propagation time $\hbar\beta$ is small, and the short-time expansion can be successfully applied.

\subsection{Time-of-Flight Graphs for BECs}

In typical BEC experiments, a trapping potential is switched off and the 
gas is allowed to expand freely during a short flight time $t$  which is of the order of several tens of miliseconds. Afterwards an absorption picture 
is taken which maps the density profile to the plane perpendicular to the laser beam. For the ideal Bose condensate, the density profile after time $t$ is given by
\begin{equation}
n(\mathbf x, t)=N_0|\psi_0(\mathbf x, t)|^2+\sum_{n\geq1}N_n |\psi_n(\mathbf x, t)|^2\, ,
\label{eq:ntpsi}
\end{equation}
where the density profile has to be integrated along the imaging axis, and the eigenstates $\psi_n(\mathbf x, t)$ are propagated according to the free Hamiltonian, containing only the kinetic term, since the 
trapping potential is switched off. If the energy eigenstates are available exactly, either analytically or numerically, their propagation time can be calculated by performing two consecutive Fourier 
transformations:
\begin{equation}
\hspace*{-2mm}
\psi_n(\mathbf x, t)=
\int \frac{d^3 \mathbf k\, d^3 \mathbf R}{(2\pi)^3}\,
e^{i[\mathbf k\cdot(\mathbf r-\mathbf R)-\omega_{\mathbf k}t]}\,\psi_n(\mathbf R)\, ,
\label{eq:psit}
\end{equation}
where the term $e^{-i\omega_{\mathbf k}t}$ accounts for a free-particle propagation in $\mathbf k$-space. In practical applications, when the energy eigenstates are calculated by a numerical diagonalization of 
space-discretized transition amplitudes, the natural way to calculate the above free-particle  time evolution is to use Fast Fourier Transform (FFT) numerical libraries. 

\begin{figure}[!t]
\centering
\includegraphics[width=3.8cm]{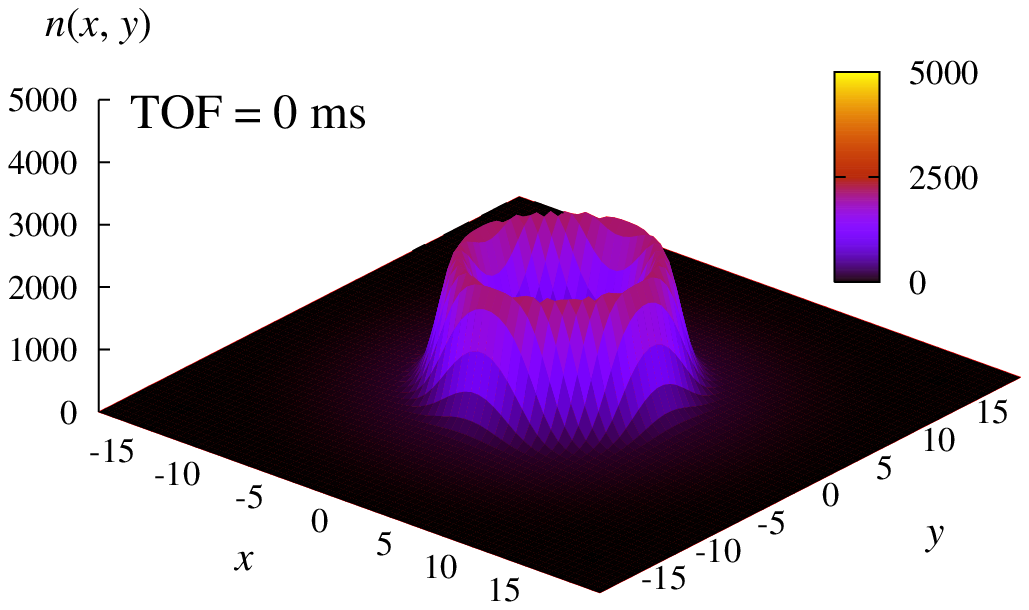}
\includegraphics[width=3.8cm]{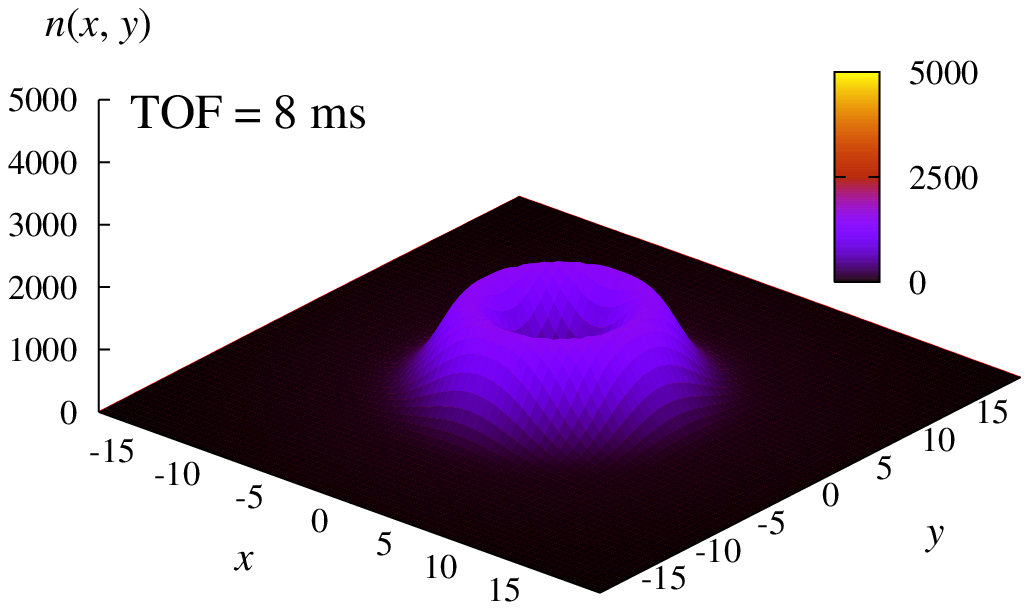}
\includegraphics[width=3.8cm]{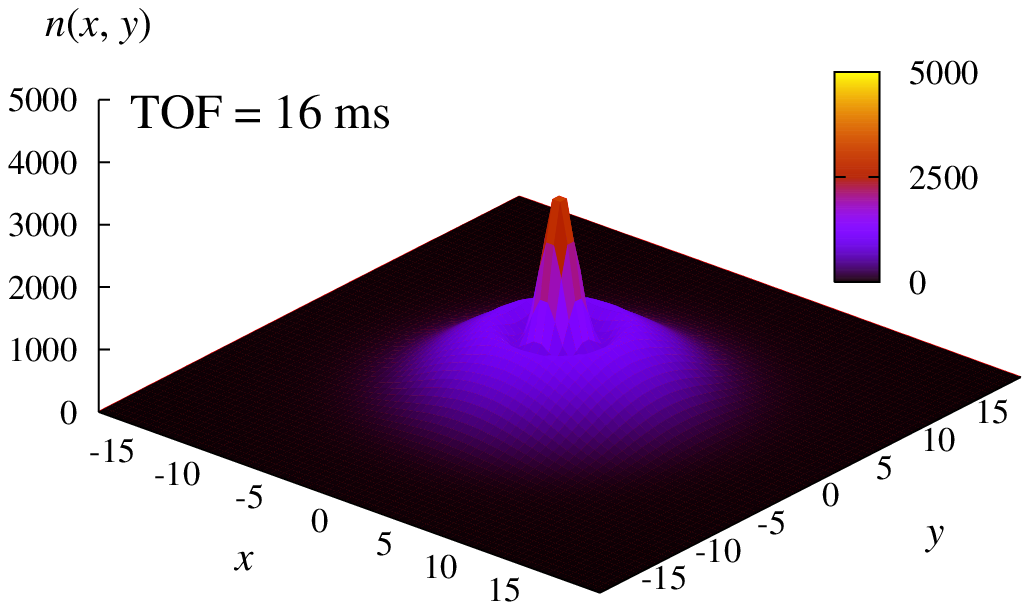}
\includegraphics[width=3.8cm]{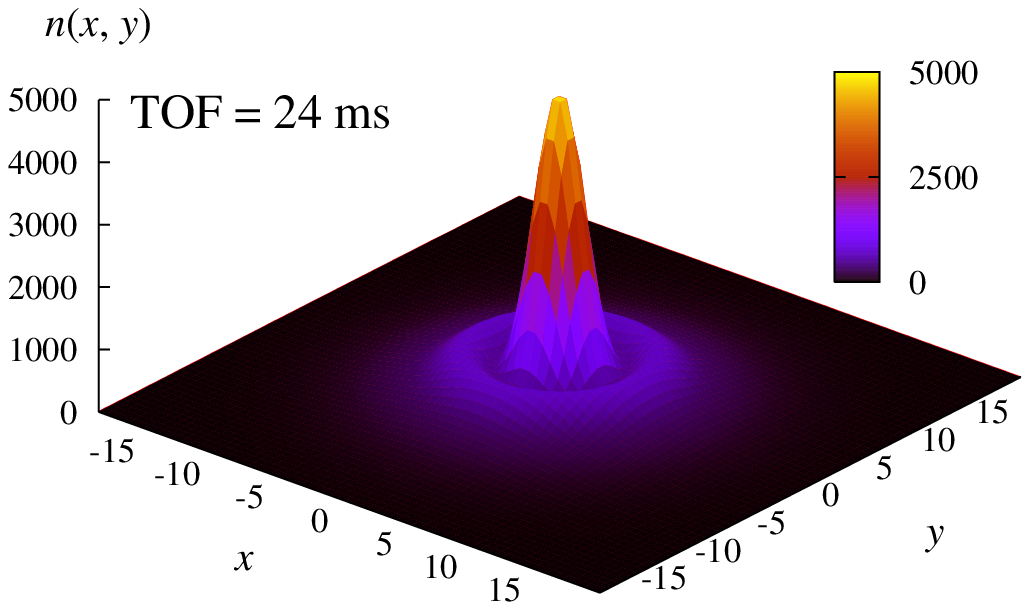}
\includegraphics[width=3.8cm]{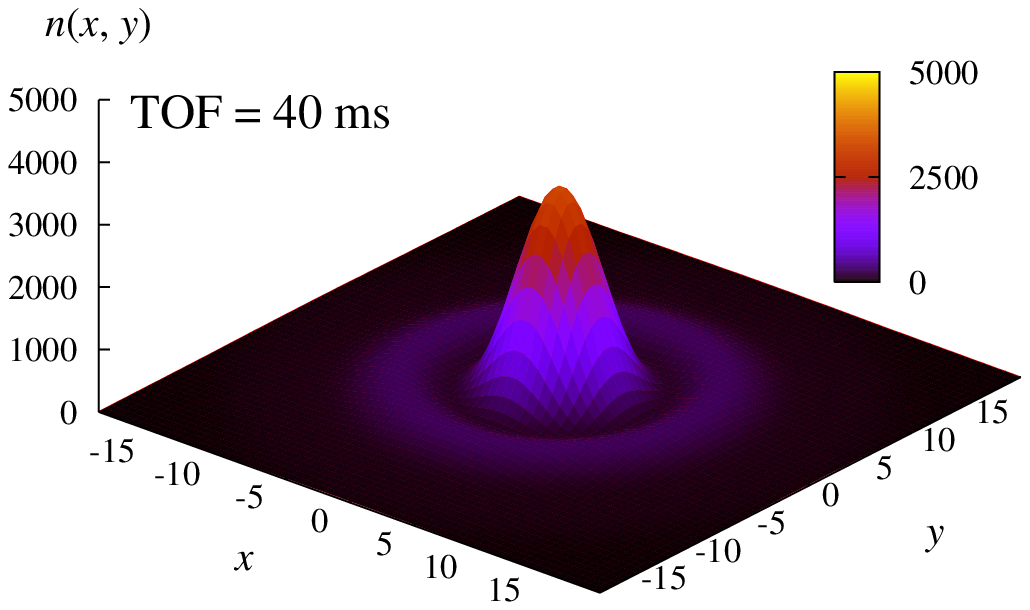}
\includegraphics[width=3.8cm]{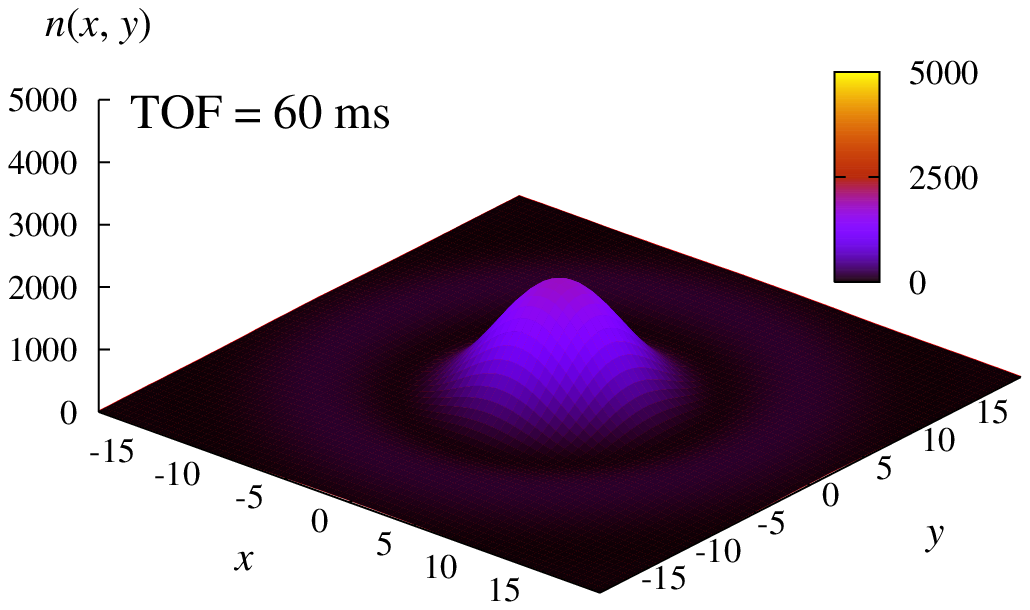}
\caption{Time-of-flight absorption density profiles in $xy$-plane for an over-critically rotating ($r=1.04$) condensate of $N=3\cdot 10^5$ atoms of $^{87}$Rb with the anharmonicity $k=k_\mathrm{BEC}$ at 
$T=30$ nK. The flight time, designated as TOF,  is given at each plot. The dimensionless unit length on all graphs corresponds to 1.34~$\mu$m and the linear size of profiles is approximately 53.6~$\mu$m. 
The discretization parameters are given in Table~\ref{tab:Emax}.}
\label{fig:tofovercrit}
\end{figure}

For high temperatures we can use a mathematically equivalent definition of the density profile which is derived again from using the cumulant expansion of occupancy numbers and 
the spectral decomposition of transition amplitudes:
\begin{eqnarray}
&&\hspace*{-12mm}n(\mathbf x, t)=N_0|\psi_0(\mathbf x, t)|^2\nonumber\\
&&\hspace*{-8mm}+\sum_{m\geq1}\left[e^{m\beta E_0}\int \frac{d^3 \mathbf k_1\, d^3 \mathbf k_2\, d^3 \mathbf X_1\, d^3 \mathbf X_2}{(2\pi)^6}\, \right.\nonumber\\
&&\hspace*{-4mm}\times e^{i[(\mathbf k_1-\mathbf k_2)\cdot\mathbf x-\mathbf k_1\cdot\mathbf X_1 + \mathbf k_2\cdot\mathbf X_2-(\omega_{\mathbf k_1}-\omega_{\mathbf k_2})t] }\nonumber\\
&&\hspace*{-4mm}\times A(\mathbf X_1, 0; \mathbf X_2, m\beta\hbar) - |\psi_0(\mathbf x, t)|^2\Bigg]\, .
\label{eq:ntA}
\end{eqnarray}
In both approaches it is first necessary to calculate the ground-state energy $E_0$ and the eigenfunction $\psi_0(\mathbf x)$, as well as the ground-state occupancy $N_0$. 
If we rely on Eqs.~(\ref{eq:ntpsi}) and 
(\ref{eq:psit}) to calculate time-of-flight graphs, we have to calculate as many eigenstates as possible by numerical diagonalization. Conversely,
if it is possible to use directly Eq.~(\ref{eq:ntA}), 
we can apply the effective action short-time expansion of thermal transition amplitudes. In both cases FFT is ideally suited for calculating time-of-flight graphs.

\subsection{Overcritical Rotation}

The case of critical and overcritical rotation $r\geq 1$ is realized in the Paris experiment by introducing the anharmonic part of the potential (\ref{eq:trap}), 
so that the condensate is confined even when the 
harmonic part of the trapping potential is completely compensated or overcompensated by the rotation. The experimental realization of this delicate balance was difficult to achieve, but nevertheless when 
the condensate was successfully confined while rotating over-critically, the measurements of its properties can be done using the standard techniques, including absorption imaging. 
Within the semiclassical 
approach one has to carefully consider this situation, since the chemical potential is defined by the minimum of the potential, and now cannot be simply set to zero anymore \cite{kling1}.
In our numerical approach, however, 
the implementation of the methods described in previous sections is straightforward even for overcritical rotation.
First one calculates energy 
eigenvalues and eigenstates using exact diagonalization, yielding negative values for the first several eigenstates. Table~\ref{tab:over} shows the resulting 
energy spectrum of an over-critically rotating condensate 
($r=1.04$) for the experimental value of the anharmonicity $k_\mathrm{BEC}$, as well as for the case of large anharmonicity $10^3~k_\mathrm{BEC}$.

\begin{figure}[!t]
\centering
\includegraphics[width=7.8cm]{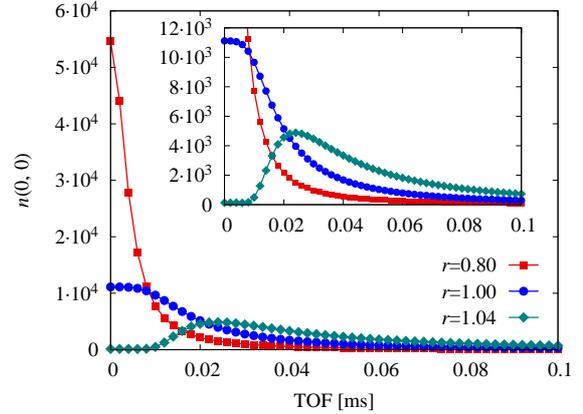}
\caption{Condensate density at the origin of $xy$-plane as a function of the time of flight (TOF) for the condensate of $N=3\cdot 10^5$ atoms of $^{87}$Rb at $T=30$ nK for several rotation frequencies 
$\Omega$ in units of $r=\Omega/\omega_\perp$. The quartic anharmonicity is $k=k_\mathrm{BEC}$ and the discretization parameters are given in Table~\ref{tab:Emax}.}
\label{fig:dorigin}
\end{figure}

Condensation temperature and other global properties as well as local properties of overcritically rotating condensates can also be calculated as before. Fig.~\ref{fig:tofovercrit} gives the 
time-of-flight absorption imaging sequence in the $xy$-plane for an
overcritically ($r=1.04$) rotating Bose-Einstein condensate with the anharmonicity $k=k_\mathrm{BEC}$ and the 
particle number $N=3\cdot 10^5$ at $T=30$ nK, exhibiting an interesting behavior. 
The initial density profile has a minimum at the origin, due to the shape of the anharmonic potential. The 
free expansion of the condensate leads to an increase in the particle density at the origin, 
and only afterwards the condensate density profile expands monotonically. Fig.~\ref{fig:dorigin} presents the time 
dependence of the 
particle density at the origin for varying rotation frequencies, parametrized by the ratio $r=\Omega/\omega_\perp$. 
We read off that approaching the critical rotation slows down the expansion of the condensate. 
For overcritical rotation this is even more pronounced, due to the appearance of the peak in the
particle density for the expansion time $t>0$. This leads to an      expansion which is typically an order of 
magnitude slower for the rotation with $r>1$.

\section{Conclusions}
\label{sec:con}
In this Letter a new method for numerically calculating  short-time transition amplitudes based on the effective action approach is applied to the study of ideal Bose gases. 
Earlier derived higher-order discretized effective actions are used for an efficient numerical calculation of both global and local properties of fast-rotating Bose-Einstein condensates.
To this end we have calculated large numbers of single-particle eigenvalues and eigenstates using an exact numerical diagonalization of the space-discretized evolution operator matrix. 
Using this information, we have calculated the condensation temperature and the ground-state occupancy of the condensate, as well as density profiles and time-of-flight absorption 
graphs. We have also shown that a critical and an overcritical rotation can be studied using the presented numerical approach, and that it leads to a substantial increase in the time 
scale for the free expansion of the condensate after the trapping potential is switched off.
Finally, we note that our
approach can also be  
used for numerical studies of properties of rotating ultra-cold Fermi gases \cite{lima}.

\section*{Acknowledgments}
This work was supported in part by the Ministry of Science and Technological Development of the Republic of Serbia, under project No. OI141035 and bilateral project PI-BEC 
funded jointly with the German Academic Exchange Service (DAAD), and the European Commission under EU Centre of Excellence grant CX-CMCS. Numerical simulations were run on 
the AEGIS e-Infrastructure, supported in part by FP7 projects EGEE-III and SEE-GRID-SCI.


\begin {thebibliography}{00}

\bibitem{pitaevskii}
L. Pitaevskii and S. Stringari, 
{\it Bose-Einstein Condensation} 
(Oxford University Press, Oxford, 2003).

\bibitem{pethick}
C. Pethick and H. Smith, 
{\it Bose-Einstein Condensation in Dilute Gases}, 
2nd edition (Cambridge University Press, Cambridge, 2008).

\bibitem{griffin}
A. Griffin, T. Nikuni, and E. Zaremba, 
{\it Bose-Condensed Gases at Finite Temperatures} 
(Cambridge University Press, Cambridge, 2009).

\bibitem{stoof}
H.~T.~C. Stoof, K.~B. Gubbels, and D.~B.~M. Dickerscheid, 
{\it Ultracold Quantum Fields} 
(Springer, Berlin, 2009).

\bibitem{fetter}
A.~L. Fetter, 
Rev. Mod. Phys. {\bf 81}, 647 (2009).

\bibitem{fetter2}
A.~L. Fetter, 
Phys. Rev. A {\bf 64}, 063608 (2001).

\bibitem{dalibard1}
V. Bretin, S. Stock, Y. Seurin, and J. Dalibard, 
Phys. Rev. Lett. {\bf 92}, 050403 (2004).

\bibitem{bloch}
I. Bloch, J. Dalibard, and W. Zwerger, 
Rev. Mod. Phys. {\bf 80},  885 (2008).

\bibitem{kling2}
S. Kling and A. Pelster,
Laser Physics {\bf 19}, 1072 (2009).

\bibitem{kling1}
S. Kling and A. Pelster,
Phys. Rev. A {\bf 76}, 023609 (2007).

\bibitem{bogojevicprl}
A. Bogojevi\' c, A. Bala\v z, and A. Beli\' c, 
Phys. Rev. Lett. {\bf 94}, 180403 (2005).

\bibitem{bogojevicprb}
A. Bogojevi\' c, A. Bala\v z, and A. Beli\' c, 
Phys. Rev. B {\bf 72}, 064302 (2005).

\bibitem{bogojevicpla}
A. Bogojevi\' c, A. Bala\v z, and A. Beli\' c, 
Phys. Lett. A {\bf 344}, 84 (2005).

\bibitem{bogojevicplamb}
A. Bogojevi\' c, I. Vidanovi\' c, A. Bala\v z, and A. Beli\' c, 
Phys. Lett. A {\bf 372}, 3341 (2008).

\bibitem{balazpre}
A. Bala\v z, A. Bogojevi\' c, I. Vidanovi\' c, and A. Pelster, 
Phys. Rev. E {\bf 79}, 036701 (2009).

\bibitem{pqseeo1}
I. Vidanovi\' c, A. Bogojevi\' c, and A. Beli\' c, 
Phys. Rev. E {\bf 80}, 066705 (2009).

\bibitem{pqseeo2}
I. Vidanovi\' c, A. Bogojevi\' c, A. Bala\v z, and A. Beli\' c,  
Phys. Rev. E {\bf 80}, 066706 (2009).

\bibitem{kleinert}
H. Kleinert,
\emph{Path Integrals in Quantum Mechanics, Sta\-tistics, Polymer Physics, and Financial Markets},
5th ed. (World Scientific, Singapore, 2009).

\bibitem{pla-danica}
D. Stojiljkovi\' c, A. Bogojevi\' c, and A. Bala\v z, 
Phys. Lett. A {\bf 360}, 205 (2006).

\bibitem{sethia} 
A. Sethia, S. Sanyal, and Y. Singh, 
J. Chem. Phys. {\bf 93}, 7268 (1990).

\bibitem{sethiacpl1} 
A. Sethia, S. Sanyal, and F. Hirata, 
Chem. Phys. Lett. {\bf 315}, 299 (1999).

\bibitem{sethiajcp} 
A. Sethia, S. Sanyal, and F. Hirata, 
J. Chem. Phys. {\bf 114}, 5097 (2001).

\bibitem{sethiacpl2} 
S. Sanyal and A. Sethia, 
Chem. Phys. Lett. {\bf 404}, 192 (2005).

\bibitem{speedup}
SPEEDUP $C$ language and Mathematica code:\\ 
{\tt http://www.scl.rs/speedup}

\bibitem{lima}
K. Howe, A. R. P. Lima, and A. Pelster:
Europ. Phys. J. D {\bf 54}, 667 (2009).

\end{thebibliography}
\end{document}